\documentclass[11pt,twoside]{atmp}

\usepackage{amsmath,amssymb}

\usepackage[all]{xy}
\usepackage{color}
\copyrightnotice{2012}{xx}{1}{19} 

\usepackage{latexsym}

\begin{document}

\title{\textbf{A conducting ball in an axial electric field}}

\arxurl{x}

\author[A.Savchenko]{\textbf{Alexander Savchenko}}

\address{Institute of Computational Mathematics and Mathematical Geophysics,
 Siberian Branch RAS, Novosibirsk, Russia}  
\addressemail{savch@ommfao1.sscc.ru}

\begin{abstract}
We describe the distribution of a charge, the electric moments of arbitrary 
order and the force acting on a conducting ball on the axis of the axial 
electric field. We determine the full charge and the dipole moments of the 
first order for a conducting ball in an arbitrary inhomogeneous harmonic 
electric field. All statements are formulated in the form of theorems with 
proofs basing on properties of the matrix of moments of the Legendre 
polynomials. The analysis and proof of these properties are presented in 
Appendix.

\end{abstract}

\maketitle

\section{\textbf{Introduction}}

When a grounded conducting ball is installed on the axis of an external 
axial harmonic field, an induced charge arises on its surface. It shields 
the initial field so that the total potential of exterior and shielded 
fields inside the ball be equal to zero. If the potential of the ball with 
radius $r$ is equal to $U$, then the distribution of the charge density on 
its surface will differ from that of a grounded ball on a constant, that is 
equal to ${\varepsilon _0 U} \mathord{\left/ {\vphantom {{\varepsilon _0 U} 
r}} \right. \kern-\nulldelimiterspace} r$, where $\varepsilon _0 $ is 
electric constant. 

\cutpage 
\setcounter{page}{2}

The density of a surface charge and the dipole moment of a ball in a 
homogeneous electric field are well known (see, for example, [1]). We 
propose the new efficient method for determining the surface charge density 
of a conducting ball in an inhomogeneous electric field. The required 
density is the solution of a one-dimensional integral equation, when the 
external field is a harmonic in the ball's volume. From here we obtain the 
analytical solution for the charge density in the case when the external 
field potential is a polynomial of arbitrary order on the axis of symmetry. 
The three theorems are proved. The first one consists of the two statements. 
The charge of the ball is fully determined by the value of potential of the 
external field at the center of the ball. The dipole moment of the ball is 
fully determined by the value of the external field intensity at the center 
of the ball. The second theorem is a generalization the first one and it 
declares that the dipole moment of $2k$ order of the ball is fully 
determined by the values of the first $k+1$ even derivatives from the 
potential at the ball's center, and the dipole moment of $2k+1$ order of the 
ball is fully determined by the values of the first $k+1$ odd derivatives 
from the potential at the ball's center. In the third theorem, we find the 
analytical form for the force acting on the ball when the axial potential of 
external field is a polynomial. In Appendix, we present and prove properties 
of the matrix of moments of the Legendre polynomials that are necessary for 
proving all the theorems. In particular, these are lemmas about linear 
dependence and orthogonality. We obtain explicit expressions for the entries 
of the inverse matrix. This circumstance allows one to find the surface 
charge density analytically. 

The investigations presented in Appendix could be of special interest in 
theory of orthogonal polynomials. By this reason, formulas in Appendix are 
given with independent double numeration.

\noindent

\section{\textbf{The potential of a conducting axially symmetric body 
on the axis of an electric field}}

The potential of the electric field $\varphi \left( \bf r \right)$, when it is 
disturbed by a conducting body, is equal to the sum $\varphi _0 \left( \bf r 
\right)+u\left( \bf r \right)$, where $\varphi _0 \left( \bf r \right)$ is the 
potential of the external initial field, and $u\left( \bf r \right)$ is the 
potential generated by surface charges of the body. The potential $u\left( \bf r 
\right)$ is determined by the formula
\begin{equation}
\label{eq1}
u\left( \bf r \right)=\frac{1}{4\pi {\kern 1pt}\varepsilon _0 }\int\limits_S 
{\frac{\sigma \left( {\bf{r}'} \right)dS}{\left| {\bf r- \bf{r}'} \right|}} ,
\end{equation}
where $\bf{r}'$ is a coordinate on the surface $S$ of the body, $\sigma \left( 
{\bf{r}'} \right)$ is the surface charge density that provides the 
equipotentiality of the surface $S$. The axially symmetric initial electric 
potential $\varphi _0 \left( {Z,X} \right)$ induces the axially symmetric 
distribution of the surface charges $\sigma \left( z \right)$ on the surface 
of a conducting axially symmetric body. These charges generate the following 
potential on the axis of the electric field at a point $s$
\begin{equation}
\label{eq2}
u\left( s \right)=\frac{1}{2\varepsilon _0 }\int\limits_{z_1 }^{z_2 } 
{\frac{h\left( z \right)}{\sqrt {\left( {z-s} \right)^2+h^2\left( z \right)} 
}\sigma \left( z \right)dl\left( z \right)} ,
\end{equation}
where $dl\left( z \right)$ is an element of length of the generator line of 
the body of revolution, $h\left( z \right)$ is a radius of the cross-section 
of the body at a distance $z$ from the origin, $z_1 $ and $z_2 $ are 
coordinates of the intersection of the axis of symmetry with the body 
surface. If the electric potential $u\left( s \right)$ compensates the 
initial potential $\varphi _0 \left( s \right)=\varphi _0 \left( {s,0} 
\right)$ on the axis, then the total potential inside the body will be equal 
to zero not only on the axis, but also on the whole volume of the body, 
because every component of the potential satisfies the Laplace equation [2].

\section{\textbf{The charges on the surface of a conducting ball}}

For a conducting ball with a radius $r$, placed on the axis in the external 
electric field, equation (2) takes the form
\begin{equation}
\label{eq3}
\int\limits_{-r}^r {\frac{r\,\sigma \left( z \right)}{\sqrt {s^2+r^2-2sz} 
}\,} dz=-2\varepsilon _0 \varphi _0 \left( s \right)\quad ,
\quad
-r<s<r\quad .
\end{equation}
Let us consider dimensionless variables $\xi =s/r$, $\eta =z/r$. Then 
equation (3) can be written down as
\begin{equation}
\label{eq4}
\int\limits_{-1}^1 {\frac{r\,\sigma \left( {r\eta } \right)}{\sqrt {\xi 
^2+1-2\xi \eta } }\,} d\eta =-2\varepsilon _0 \varphi _0 \left( {r\xi } 
\right)\quad ,
\quad
-1<\xi <1.
\end{equation}
If $\sigma \left( {r\eta } \right)=\eta ^n$, where $n$ is a natural number, 
then, integrating the left-hand side in (4), we obtain, using the 
mathematical induction, a polynomial of $n$ degree also in the right-hand 
side depending on the variable $\xi $. Therefore when the right-hand side in 
equation (4) varies along the axis as a polynomial of $n$ degree
\[
-\varphi _0 \left( {r\xi } \right)=\sum\limits_{i=1}^{n+1} {{b}'_i \xi 
^{i-1}} ,
\]
then $\sigma (r\eta )$ is also a polynomial of $n$ degree 
\[
\sigma (r\eta )=\frac{2\varepsilon _0 }{r}\sum\limits_{j=1}^{n+1} {{c}'_j 
\eta ^{j-1}} ,
\]
with the coefficients ${c}'_j $ that can be determined from the equation
\[
\bf F \bf{c}'= \bf{b}',
\]
where ${\bf{c}'}=({c}'_1 ,\ldots ,{c}'_{n+1} )^T, \quad \quad {\bf{b}'}=({b}'_1 ,\ldots 
,{b}'_{n+1} )^T,\quad  \quad {\bf F}=\left\{ {\,F_{ij} } \right\},\quad $ 
$i,j=1,\ldots ,n+1$, and the matrix $\bf F$ is an upper triangular one. The 
nonzero entries in this matrix are equal to
\begin{equation}
\label{eq5}
F_{ij} =\frac{1}{\left( {i-1} \right)!}\int\limits_{-1}^1 {\left( 
{\frac{\partial ^{i-1}}{\partial \xi ^{i-1}}\frac{1}{\sqrt {\xi ^2+1-2\xi 
\eta } }} \right)_{\xi =0} \eta ^{j-1}d\eta } \,\quad .
\end{equation}
Then the matrix ${\bf G=F^{-1}}=\left\{ {G_{ij} } \right\}$ will also be the upper 
triangular one. 

Let the potential of an external electric field vary along the axis as a 
polynomial of $n$ degree
\begin{equation}
\label{eq6}
-\varphi _0 \left( s \right)=\sum\limits_{i=1}^{n+1} {b_i s^{i-1}} .
\end{equation}
Then the density of a surface charge is a polynomial of the same degree
\begin{equation}
\label{eq7}
\sigma (z)=\frac{2\varepsilon _0 }{r}\sum\limits_{j=1}^{n+1} {c_j z^{j-1}} 
,
\end{equation}
and the coefficients can be determined from the equation
\begin{equation}
\label{eq8}
\bf F R c=R b\quad .
\end{equation}
Hence, the required coefficients can be found in the explicit form
\begin{equation}
\label{eq9}
\bf c= R^{-1} G R b
\end{equation}
or
\begin{equation}
\label{eq10}
\bf c=\tilde{G} b, 
\end{equation}
where ${\bf c}=(c_1 ,\ldots ,c_{n+1} )^T$, ${\bf b}=(b_1 ,\ldots ,b_{n+1} )^T$, 
${\bf R}=diag\left\{ {1,r,\ldots ,r^n} \right\}$, and entries of the matrix 
$\bf \tilde{G}$ are determined by a simple equality $\tilde {G}_{ij} 
=r^{j-i}G_{ij} $, $i,j=1,\ldots ,n+1$. 

The explicit form of entries of the matrix $\bf G$ is found in Appendix 
(Property 8). Hence, the density of a surface charge in the explicit form 
can be found, when the potential of an external electric field on the axis 
is a polynomial of an arbitrary degree.

\section{\textbf{The full charge and dipole moment of a conducting ball}}

The full charge $Q$ and dipole moment $\mathcal D$ of a conducting ball with 
radius $r$ on the axis of the axially symmetric electric field are 
determined by the formulas
\begin{equation}
\label{eq11}
Q=2\pi \int\limits_{-r}^r {\sigma \left( z \right)} {\kern 1pt}h\left( z 
\right){\kern 1pt}dl\left( z \right)=2\pi r\int\limits_{-r}^r {\sigma \left( 
z \right){\kern 1pt}dz} ,
\end{equation}
\begin{equation}
\label{eq12}
{\mathcal D}=2\pi r\int\limits_{-r}^r {z\sigma \left( z \right)} {\kern 1pt}dz.
\end{equation}
If the potential of the external field is a linear function on the axis of 
symmetry $-\varphi _0 \left( s \right)=b_1 +b_2 s$, then from (7) and (10) 
we obtain
\begin{equation}
\label{eq13}
\sigma \left( z \right)=\frac{2\varepsilon _0 }{r}\left( {c_1 +c_2 z} 
\right)=\frac{2\varepsilon _0 }{r}\left( {G_{11} b_1 +G_{22} b_2 z} 
\right).
\end{equation}
From Property 8 (Appendix), which determines the entries of the inverse 
matrix $\bf G$, one can easily obtain $G_{11} =\frac{1}{2}$, $G_{22} 
=\frac{3}{2}$. Thus, from formulas (11)-(13) we obtain
\begin{equation}
\label{eq14}
Q=4\pi \varepsilon _0 rb_1 \quad ,
\end{equation}
\begin{equation}
\label{eq15}
{\mathcal D}=4\pi \varepsilon _0 r^3b_2 \quad .
\end{equation}
\newtheorem{teo}{Theorem}
\begin{teo}
\textbf{The full charge and dipole moment of a 
conducting ball are determined by formulas (14) and (15) in any external 
axially symmetric field that is harmonic inside the ball.}
\end{teo}
\textit{Proof. }Theorem is proved by induction. Assume that formulas (14) and 
(15) are valid 
provided that the potential $\varphi _0 \left( s \right)$ is a polynomial of 
$n-1$ degree. This means that such equalities are valid for polynomials 
$\sigma \left( z \right)$ of $n-1$ degree, with coefficients that are 
determined from the solution of a system of linear equations of order $n$ 
with an arbitrary right-hand side, where entries of the matrix $\bf F$ are 
determined by equalities (5). We prove the validity of equalities (14) and 
(15), when the potential $\varphi _0 \left( s \right)$, determined in (6), 
is a polynomial of $n$ degree. In this case, the coefficients of polynomial 
(7) are determined from the system of linear equations (8) of order $n+1$, 
in which the matrix $\bf F$ is the upper triangular one. Then the last component 
of the vector $\bf c$ can be determined immediately from the equality
\begin{equation}
\label{eq16}
c_{n+1} =\frac{b_{n+1} }{F_{n+1,n+1} },
\end{equation}
and the system of equations (8) can be written down as a system of linear 
equations of order $n$ 
\begin{equation}
\label{eq17}
\begin{array}{l}
 {\bf F}^{\left( n \right)}{\bf R}^{\left( {n-1} \right)}{\bf c}^{\left( {n-1} 
\right)}={\bf R}^{\left( {n-1} \right)}{\bf b}^{\left( {n-1} \right)}-c_{n+1} 
r^n{\bf f}_{\bullet ,n+1} = \\ 
 {\bf R}^{\left( {n-1} \right)}\left[ {{\bf b}^{\left( {n-1} \right)}-c_{n+1} 
r^n({\bf R}^{(n-1)})^{-1}{\bf f}_{\bullet ,n+1} } \right] \\ 
 \end{array}\quad ,
\end{equation}
where

${\bf F}^{\left( n \right)}=\left\{ {F_{ij} } \right\},
\quad
i,j=1,\ldots ,n;
\quad
{\bf c}^{\left( {n-1} \right)}=\left( {c_1 ,c_2 ,\ldots ,c_n } \right)^T;
\quad
{\bf b}^{\left( {n-1} \right)}=\left( {b_1 ,b_2 ,\ldots ,b_n } \right)^T;
\quad
{\bf R}^{\left( {n-1} \right)}=diag\left\{ {1,r,\ldots ,r^{n-1}} \right\};
\quad
{\bf f}_{\bullet ,n+1} =\left( {F_{1,n+1} ,\ldots ,F_{n,n+1} } \right)^T,$ and 
$c_{n+1} $ is not just a required variable, but it is determined by entries 
of the matrix $\bf F$ and the right-hand side $\bf b$.

Denote $\sigma ^{\left( {n-1} \right)}\left( z \right)=\frac{2\varepsilon _0 
}{r}\sum\limits_{j=1}^n {c_j z^{j-1}} $ and write equations (11) and (12) in 
the form
\begin{equation}
\label{eq18}
Q=Q^{\left( {n-1} \right)}+4\pi \varepsilon _0 c_{n+1} 
r^{n+1}\int\limits_{-1}^1 {z^ndz} ,
\end{equation}
\begin{equation}
\label{eq19}
{\mathcal D}={\mathcal D}^{\left( {n-1} \right)}+4\pi \varepsilon _0 c_{n+1} 
r^{n+2}\int\limits_{-1}^1 {z^{n+1}dz} ,
\end{equation}
where 
\[
Q^{\left( {n-1} \right)}=2\pi r\int\limits_{-r}^r {\sigma ^{\left( {n-1} 
\right)}\left( z \right)} \,dz\quad ,
\]
\[
{\mathcal D}^{\left( {n-1} \right)}=2\pi r\int\limits_{-r}^r {z\sigma ^{\left( 
{n-1} \right)}\left( z \right)\,dz} .
\]
Since the induction assumption is valid for the system of equations (17) of 
order $n$, then 
\begin{equation}
\label{eq20}
Q^{\left( {n-1} \right)}=4\pi \varepsilon _0 r\left( {b_1 -c_{n+1} 
r^nF_{1,n+1} } \right),
\end{equation}
\begin{equation}
\label{eq21}
{\mathcal D}^{\left( {n-1} \right)}=4\pi \varepsilon _0 r^3\left( {b_2 -c_{n+1} 
r^{n-1}F_{2,n+1} } \right).
\end{equation}
From (5) we have
\begin{equation}
\label{eq22}
F_{1,n+1} =\int\limits_{-1}^1 {\eta ^nd\eta } ,
\end{equation}
\begin{equation}
\label{eq23}
F_{2,n+1} =\int\limits_{-1}^1 {\left( {\frac{\partial }{\partial 
s}\frac{1}{\sqrt {1+s^2-2s\eta } }} \right)} _{s=0} \eta ^nd\eta 
=\int\limits_{-1}^1 {\eta ^{n+1}d\eta } .
\end{equation}
The proof of the theorem follows from formulas (18)-(23). 
\newtheorem{cor}{Corollary}
\begin{cor}
\textbf{Formulas (14) and (15) are valid for any 
external harmonic fields that are not necessarily axially symmetric 
fields.}
\end{cor}
\textit{Proof. }The electric field, generated by a point charge is always axially symmetric 
for the ball, when the axis passes through its center and this charge. Then 
any distribution of point charges excites axial fields with axes passing 
through the center of the ball. Since Theorem 1 is also valid for the sum of 
electric fields with axes passing through the center of the ball and, since 
an electric field can always be presented as superposition of point charges, 
then Theorem 1 is valid for any electric field. 

\section{\textbf{The dipole moments of higher orders of a conducting ball}}

Now let us discuss the deduction of the formula for the dipole moment of a 
conducting ball of $m$ order, which is determined by the formula
\[
{\mathcal D}_m =2\pi r\int\limits_{-r}^r {z^m\sigma \left( z \right)dz} .
\]
\begin{teo}
\textbf{The dipole moment of the ball of 
 $m$ \textbf{\textit{order}}\textbf{, 
}\textbf{\textit{when the potential of the external field }}$\varphi _0 
\left( s \right)$\textbf{\textit{on the axis of symmetry is a polynomial of 
}}$n$\textbf{\textit{ degree is equal to}}
\begin{equation}
\label{eq24}
{\mathcal D}_m =2\pi \varepsilon _0 r^{m+1}\sum\limits_{i=\delta ,\left( 2 
\right)}^{m+1} {\left( {2i-1} \right)} \,r^{i-1}F_{i,m+1} b_i \quad ,
\end{equation}
\textbf{\textit{where }}$\delta =1$\textbf{\textit{ if }}$m$\textbf{\textit{ 
is even and }}$\delta =2$\textbf{\textit{ if }}$m$\textbf{\textit{ is odd.}}}
\end{teo}
In formula (24) and further the index ``(2)'' in the lower part of the sum 
symbol means that summation is carried out with a step equal to 2. Thus, in 
this case, the summation can be done only by odd or only by even values of 
the index $i$, depending on its first value.

\textit{Proof}.

From (7) and Theorem 1 follows 
\[
{\mathcal D}_m =2\pi r\int\limits_{-r}^r {z^m\sigma \left( z \right)dz=4\pi 
\varepsilon _0 \int\limits_{-r}^r {\sum\limits_{j=1}^{n+1} {c_j 
z^{m+j-1}dz=4\pi \varepsilon _0 r{b}'_1 } } } ,
\]
where ${b}'_1 $ is the first component of the right-hand side in the system 
of equations
\begin{equation}
\label{eq25}
\left( {\begin{array}{l}
 F_{11} \quad 0\quad F_{13} \quad 0 \ldots \,\,F_{1,M} \\ 
\: 0\quad \;F_{22} \;\; 0\:\quad F_{24} \ldots \,F_{2,M} \\ 
\: 0\quad \;\;\, 0\quad F_{33} \quad 0\ldots\;\, F_{3,M} \\
\: 0\quad \;\;\, 0 \;\;\quad 0 \quad F_{44} \ldots F_{4,M} \\ 
\: \ldots \;\; \ldots \;\; \ldots \;\;\ldots \;\;\ldots \;\; \ldots \\ 
\: 0\quad \;\;\, 0 \;\;\quad 0\quad\;0\,\ldots \,F_{M,M} \\ 
 \end{array}} \right)\;\left( {\begin{array}{l}
 \left. {\begin{array}{l}
 \quad 0 \\ 
 \vdots \\ 
 \quad 0 \\ 
 \end{array}} \right\} m \\ 
 \;\;\;r^mc_1 \\ 
 \quad \vdots \\ 
 r^{M-1}c_{n+1} \\ 
 \end{array}} \right)=\left( {\begin{array}{l}
 \quad {b}'_1 \\ 
 \quad r{b}'_2 \\ 
 \quad r^2{b}'_3 \\ 
 \;\quad \;\,\vdots \\ 
 \;\;\,\quad \vdots \\ 
 r^{M-1}{b}'_{M} \\ 
 \end{array}} \right),
\end{equation}
where $M=n+m+1$.
From (25) follows 
\begin{equation}
\label{eq26}
\bf \tilde {F}\tilde {R}c=R{b}'\quad ,
\end{equation}
where ${\bf{b}'}=\left( {{b}'_1 ,\ldots ,{b}'_{n+1} } \right)^T$, ${\bf \tilde 
{R}}=diag\left\{ {r^m,r^{m+1},\ldots ,r^{M-1}} \right\}$,
\[
{\bf\tilde {F}}=\left( {\begin{array}{l}
 F_{1,m+1} \quad\; F_{1,m+2} \quad \ldots \quad F_{1,M} \\ 
 F_{2,m+1} \quad\; F_{2,m+2} \;\;\;\:\ldots \quad F_{2,M} \\ 
 \;\;\ldots \quad \quad \quad\ldots \quad \quad\;\ldots \quad \;\;\ldots \\ 
 F_{n+1,m+1} \;F_{n+1,m+2} \;\,\ldots \:F_{n+1,M} \\ 
 \end{array}} \right).
\]
We express the vector $\bf {b}'$ through the vector $\bf b$, using (9): 
\[
\bf \tilde {F}\tilde {R}R^{-1}GRb=R{b}'.
\]
Consequently,
\begin{equation}
\label{eq27}
r^m{\bf \tilde {F}GRb=R{b}'}\quad .
\end{equation}
Hence, the first component of the vector $\bf {b}'$ is equal to the first line 
of the matrix $\bf L=\tilde {F}G$, multiplied by the vector $\bf Rb$ and the number 
$r^m$. We will prove that only $L_{1,\delta } ,L_{1,\delta +2} ,\ldots 
,L_{1,m+1} $ are non-zero entries in this line.

Because the matrix $\bf G$ is inverse to the matrix $\bf F$, the vectors ${\bf g}_{\bullet 
,j} $ if $j=m+2,\ldots ,n$, are orthogonal to the vectors ${\bf f}_{i,\bullet } $ 
when $i=1,\ldots ,m+1$. From Lemma about a linear combination of vectors 
(see Appendix) it follows that the vector ${\bf {f}'}=\left( {F_{1,m+1} ,F_{1,m+2} 
,\ldots ,F_{1,M} } \right)^T$is a linear combination of the vectors 
${\bf f}_{\delta ,\bullet } ,{\bf f}_{\delta +2,\bullet } \ldots ,{\bf f}_{m+1,\bullet } 
$ and, hence, it will be orthogonal to all the vectors ${\bf g}_{\bullet 
,j} $, $j=m+2,\ldots ,n$. Hence, $L_{1,l} =0$ when $l>m+1$.

Let us find the first $m+1$ entries in the first line of the matrix $\bf L$. 
Because the vector $\bf {f}'$ is a linear combination of the vectors ${\bf f}_{\delta 
,\bullet } ,{\bf f}_{\delta +2,\bullet } \ldots ,{\bf f}_{m+1,\bullet } $ with 
the coefficients $\alpha _\delta ,\alpha _{\delta +2} ,\ldots ,\alpha _{m+1} 
$, then, by virtue of orthogonality of the vectors ${\bf f}_{i,\bullet } $ and 
${\bf g}_{\bullet ,l} $, when $i\ne l$ and $l\le m+1$, we have
\[
L_{1l} =\left( {{\bf {f}'},{\bf g}_{\bullet ,l} } \right)=\sum\limits_{i=\delta ,\left( 
2 \right)}^{m+1} {\alpha _i \left( {{\bf f}_{i,\bullet } ,{\bf g}_{\bullet ,l} } 
\right)=\alpha _l } ,
\]
where $\alpha _l =\frac{2l-1}{2}F_{l,m+1} $ (see formula (9.1) in Appendix). 
Then from (27) follows 
\[
{b}'_1 =\frac{r^m}{2}\sum\limits_{i=\delta ,\left( 2 
\right)}^{m+1} {\left( {2i-1} \right)r^{i-1}F_{i,m+1} b_i } ,
\]
and 
\[
{\mathcal D}_m =2\pi \varepsilon _0 r^{m+1}\sum\limits_{i=\delta ,\left( 2 
\right)}^{m+1} {\left( {2i-1} \right)r^{i-1}F_{i,m+1} b_i } .
\] 

\section{\textbf{The force acting on a conducting ball}}

The force ${\mathcal F}$, acting on the conductive ball with a radius $r$ that is 
on the axis of the axially symmetric electric field, is equal to
\[
{\mathcal F}=2\pi \int\limits_{-r}^r {p\left( z \right)} {\kern 1pt}h\left( z 
\right){\kern 1pt}dh\left( z \right),
\]
where $p\left( z \right)$ is electric pressure that is determined by the 
formula $p\left( z \right)=-\frac{\sigma ^2\left( z \right)}{2\varepsilon _0 
}$ [1]. From here it follows that
\begin{equation}
\label{eq28}
{\mathcal F}=\frac{\pi }{\varepsilon _0 }\int\limits_{-r}^r {z\sigma ^2\left( z 
\right)dz} \quad .
\end{equation}
\begin{teo}
\textbf{The force ${\mathcal F}$, 
acting on the conducting ball with a radius $r$, 
located on the axis of the axially symmetric electric field 
with a potential varying along the axis as polynomial of 
$n$ degree (see formula (\ref{eq6})), is uniquely defined by 
coefficients of this polynomial $b_1 ,\ldots ,b_{n+1} $ 
and is equal to
\[
{\mathcal F}=4\pi \varepsilon _0 \sum\limits_{i=1}^n {i\,r^{2i-1}b_i b_{i+1} } .
\]}
\end{teo}
\textit{Proof}. The theorem is proved by induction.

Let $n=1$. Then from (\ref{eq7}) and (10) follows
\[
\sigma \left( z \right)=\frac{2\varepsilon _0 }{r}\left( {c_1 +c_2 z} 
\right)=\frac{2\varepsilon _0 }{r}\left( {G_{11} b_1 +G_{22} b_2 z} 
\right).
\]
From Property 8 (Appendix), which determines the entries of the inverse 
matrix $\bf G$, it follows that $G_{11} =\frac{1}{2}$, $G_{22} =\frac{3}{2}$. 
Then
\[
{\mathcal F}=\frac{4\pi \varepsilon _0 }{r^2}\int\limits_{-r}^r {z\left( 
{\frac{1}{2}b_1 +\frac{3}{2}b_2 z} \right)^2dz=4\pi \varepsilon _0 rb_1 b_2 
} .
\]
Assume that the statement of the theorem is valid for $k=n-1$ and we will 
prove its validity for $k=n$.

Since the integral of an odd function from $-r$ to $r$ equals zero, then
\begin{equation}
\label{eq29}
\begin{array}{l}
 \int\limits_{-r}^r z \left( {\sum\limits_{j=1}^{n+1} {c_j z^{j-1}} } 
\right)^2dz= \\ 
 \int\limits_{-r}^r z \left( {\sum\limits_{j=1}^n {c_j z^{j-1}} } 
\right)^2dz+\;2c_{n+1} \int\limits_{-r}^r {z^{n+1}} \left( 
{\sum\limits_{j=1}^n {c_j z^{j-1}} } \right)dz \\ 
 \end{array}\quad .
\end{equation}
From Theorem 2 it follows that
\begin{equation}
\label{eq30}
\begin{array}{l}
 \int\limits_{-r}^r {z^{n+1}} \left( {\sum\limits_{j=1}^n {c_j z^{j-1}} } 
\right)dz=\int\limits_{-r}^r {z^{n+1}} \left( {\sum\limits_{j=1}^{n+1} {c_j 
z^{j-1}} } \right)dz= \\ 
 \frac{r^{n+2}}{2}\sum\limits_{i=\delta ,\left( 2 \right)}^n {\left( {2i-1} 
\right)r^{i-1}F_{i,n+2} b_i } \\ 
 \end{array}.
\end{equation}
Let us consider the first summand in the right-hand side of (29). As in the 
proof of Theorem 1, we note that the system of equations (8) of order $n+1$ 
is equivalent to the system of equations (17) of order $n$, where the last 
component of the vector $\bf c$ is determined in (16). Because the $i$-th 
component of the vector in the square brackets in (17) equals ${b}'_i =b_i 
-c_{n+1} r^{n-i+1}F_{i,n+1} $, then, by the induction hypothesis, 
\[
\frac{\pi }{\varepsilon _0 }\int\limits_{-r}^r {z\sigma _{n-1}^2 \left( z 
\right)dz=} 4\pi \varepsilon _0 \sum\limits_{i=1}^{n-1} {ir^{2i-1}{b}'_i 
{b}'_{i+1} } ,
\]
where $\sigma _{n-1} \left( z \right)=\frac{2\varepsilon _0 
}{r}\sum\limits_{j=1}^n {c_j z^{j-1}} $. In view of the fact that $F_{i,n+1} 
F_{i+1,n+1} =0$ for any natural numbers $i$ and $n$ (Property 2, Appendix)
\begin{equation}
\label{eq31}
\begin{array}{l}
 \int\limits_{-r}^r {z\left( {\sum\limits_{j=1}^n {c_j z^{j-1}} } \right)} 
^2dz= \\ 
 \sum\limits_{i=1}^{n-1} {ir^{2i+1}\left( {b_i -c_{n+1} r^{n-i+1}F_{i,n+1} } 
\right)\left( {b_{i+1} -c_{n+1} r^{n-i}F_{i+1,n+1} } \right)} = \\ 
 \sum\limits_{i=1}^{n-1} {ir^{2i+1}b_i b_{i+1} -\sum\limits_{i=2}^n {\left( 
{i-1} \right)c_{n+1} r^{n+i+1}} } F_{i-1,n+1} b_i -\sum\limits_{i=1}^{n-1} 
{ic_{n+1} r^{n+i+1}F_{i+1,n+1} b_i } \;. \\ 
 \end{array}
\end{equation}
We continue the proof of the induction hypothesis for $k=n$, considering the 
cases of odd and even values of $n$, separately.

   1.  Let $n$ be an odd number . Then by virtue of (29) - (31), we have
\begin{equation}
\label{eq32}
\int\limits_{-r}^r {z\left( {\sum\limits_{j=1}^{n+1} {c_j z^{j-1}} } 
\right)^2dz=\sum\limits_{i=1}^{n-1} {ir^{2i+1}b_i b_{i+1} +S_n^{\left( 1 
\right)} } } ,
\end{equation}
where 

$\begin{array}{l}
 {S_n^{\left( 1 \right)} } \mathord{\left/ {\vphantom {{S_n^{\left( 1 
\right)} } {c_{n+1} }}} \right. \kern-\nulldelimiterspace} {c_{n+1} 
}=r^{n+2}\sum\limits_{i=1,\left( 2 \right)}^n {\left( {2i-1} 
\right)r^{i-1}F_{i,n+2} b_i } - \\ 
 -\sum\limits_{i=3,\left( 2 \right)}^n {\left( {i-1} 
\right)r^{n+i+1}F_{i-1,n+1} b_i } -\sum\limits_{i=1,\left( 2 \right)}^{n-2} 
{ir^{n+i+1}F_{i+1,n+1} b_i } \\ 
 \end{array}$ .

From here it follows that 
\begin{equation}
\label{eq33}
\begin{array}{l}
 {S_n^{\left( 1 \right)} } \mathord{\left/ {\vphantom {{S_n^{\left( 1 
\right)} } {c_{n+1} }}} \right. \kern-\nulldelimiterspace} {c_{n+1} }= \\ 
 r^{n+2}\left\{ {\left( {F_{1,n+2} -F_{2,n+1} } \right)b_1 +\left[ {\left( 
{2n-1} \right)F_{n,n+2} -\left( {n-1} \right)F_{n-1,n+1} } \right]r^{n-1}b_n 
} \right\}+ \\ 
 \sum\limits_{i=3,\left( 2 \right)}^{n-2} {r^{n+i+1}b_i \left[ {\left( 
{2i-1} \right)F_{i,n+2} -\left( {i-1} \right)F_{i-1,n+1} -iF_{i+1,n+1} } 
\right]\quad } . \\ 
 \end{array}
\end{equation}
Since $F_{2,n+1} =F_{1,n+2} $ for any values of $n$ (Property 5, Appendix), 
then the first summand in the right-hand side of (33) is equal to zero. From 
the recurrence relation for entries of the matrix $\bf F$ (Property 3, Appendix), 
we have
\begin{equation}
\label{eq34}
F_{i+1,n+1} =\frac{2i-1}{i}F_{i,n+2} -\frac{i-1}{i}F_{i-1,n+1} .
\end{equation}
Hence, the third summand in the right-hand side of (33) is also equal to 
zero. Since the last component of the vector $\bf c$ can be found from (16), 
then 
\[
S_n^{\left( 1 \right)} =\frac{b_n b_{n+1} }{F_{n+1,n+1} }r^{2n+1}\left[ 
{\left( {2n-1} \right)F_{n,n+2} -\left( {n-1} \right)F_{n-1,n+1} } \right].
\]
Using Lemma 1 and Lemma 2 from Appendix, we obtain
\[
\begin{array}{l}
 S_n^{\left( 1 \right)} =r^{2n+1}b_n b_{n+1} \left[ {\frac{2^{n+1}\left( 
{2n-1} \right)\left[ {\left( {n+1} \right)!} \right]^2}{\left( {2n+2} 
\right)!}-\frac{2^n\left( {n-1} \right)\left[ {\left( {n!} \right)} 
\right]^2}{\left( {2n} \right)!}} \right]\frac{\left( {2n+2} 
\right)!}{2^{n+2}n!\left( {n+1} \right)!}= \\ 
 \frac{r^{2n+1}b_n b_{n+1} }{2}\left[ {\left( {2n-1} \right)\left( {n+1} 
\right)-\left( {n-1} \right)\left( {2n+1} \right)} \right]=nr^{2n+1}b_n 
b_{n+1} \quad . \\ 
 \end{array}
\]
Thus, from (32) we have
\begin{equation}
\label{eq35}
\int\limits_{-r}^r {z\left( {\sum\limits_{j=1}^{n+1} {c_j z^{j-1}} } 
\right)^2dz=\sum\limits_{i=1}^n {ir^{2i+1}b_i b_{i+1} } } \quad .
\end{equation}

   2.  Let the number $n$ be even. Then from (29) - (31) we obtain
\[
\int\limits_{-r}^r {z\left( {\sum\limits_{j=1}^{n+1} {c_j z^{j-1}} } 
\right)^2dz=\sum\limits_{i=1}^{n-1} {ir^{2i+1}b_i b_{i+1} +S_n^{\left( 2 
\right)} } } ,
\]
where 
\[
\begin{array}{l}
 {S_n^{\left( 2 \right)} } \mathord{\left/ {\vphantom {{S_n^{\left( 2 
\right)} } {c_{n+1} }}} \right. \kern-\nulldelimiterspace} {c_{n+1} 
}=r^{n+2}\sum\limits_{i=2,\left( 2 \right)}^n {\left( {2i-1} 
\right)r^{i-1}F_{i,n+2} b_i } - \\ 
 \sum\limits_{i=2,\left( 2 \right)}^n {\left( {i-1} 
\right)r^{n+i+1}F_{i-1,n+1} b_i } -\sum\limits_{i=2,\left( 2 \right)}^{n-2} 
{ir^{n+i+1}F_{i+1,n+1} b_i } \;. \\ 
 \end{array}
\]
From here follows  
\[
\begin{array}{l}
 {S_n^{\left( 2 \right)} } \mathord{\left/ {\vphantom {{S_n^{\left( 2 
\right)} } {c_{n+1} }}} \right. \kern-\nulldelimiterspace} {c_{n+1} 
}=r^{2n+1}\left\{ {\left[ {\left( {2n-1} \right)F_{n,n+2} -\left( {n-1} 
\right)F_{n-1,n+1} } \right]b_n } \right\}+ \\ 
 \sum\limits_{i=2,\left( 2 \right)}^{n-2} {r^{n+i+1}b_i \left[ {\left( 
{2i-1} \right)F_{i,n+2} -\left( {i-1} \right)F_{i-1,n+1} -iF_{i+1,n+1} } 
\right]} \quad . \\ 
 \end{array}
\]
In view of (34), the second summand in the right-hand side of the equality 
is equal to zero and is easy to see that in this case $S_n^{\left( 2 
\right)} =S_n^{\left( 1 \right)} $. Thus, equation (35) is valid for any 
values of the index $n$. Then from (7), (28) and (35) we obtain
\[
{\mathcal F}=\frac{\pi }{\varepsilon _0 }\frac{4\varepsilon _0^2 
}{r^2}\int\limits_{-r}^r {z\left( {\sum\limits_{j=1}^{n+1} {c_j z^{j-1}} } 
\right)^2dz=} 4\pi \varepsilon _0 \sum\limits_{i=1}^n {i\,r^{2i-1}b_i 
b_{i+1} } ,
\]
and the theorem is proved.

\begin{appendix}

\begin{flushright}
\textbf{APPENDIX}
\end{flushright}

\begin{center}
\textbf{Properties of the matrix of moments of the Legendre polynomials}
\end{center}

\subsection{\textbf{The matrix $\bf F$ is the matrix of moments of the Legendre 
polynomials}}

Indeed, as the function $\frac{1}{\sqrt {1-2\xi \eta +\xi ^2} }$ is the 
generating function for the Legendre polynomials $P_n \left( \eta \right)$, 
i.e. $\sum\limits_{n=0}^\infty {P_n \left( \eta \right)} \;\xi 
^n=\frac{1}{\sqrt {1-2\xi \eta +\xi ^2} }$, then
\[
\left( {\frac{\partial }{\partial \xi ^{i-1}}\frac{1}{\sqrt {1-2\xi \eta 
+\xi ^2} }} \right)_{\xi =0} =\left( {i-1} \right)!P_{i-1} \left( \eta 
\right)
\]
for any natural values $i$. Hence,

$$F_{ij} =\int\limits_{-1}^1 {P_{i-1} \left( \eta \right)} \,\eta ^{j-1}d\eta 
\quad .
\eqno(1.1)$$

\subsection{\textbf{The odd upper diagonals of the matrix }$\bf F$ \textbf{consist 
of zero entries, i.e. }$F_{ij} =0$\textbf{ if }$i+j$\textbf{ is an odd number}}

Obviously, the polynomial $P_n \left( \eta \right)$ is even if the number $n$ 
is even, and is odd otherwise. This statement is easily proved by induction 
with allowance for the recurrence formula for the Legendre polynomials

$$P_{n+1} \left( \eta \right)=\frac{2n+1}{n+1}\eta P_n \left( \eta 
\right)-\frac{n}{n+1}P_{n-1} \left( \eta \right).
\eqno(2.1)
$$

Since the integral of an odd function is equal to zero, then Property 2 is 
proved.

\subsection{\textbf{The entries of the matrix} $\bf F$\textbf{ satisfy the 
recurrence relation}}

$$F_{ij} =\frac{2i-3}{i-1}F_{i-1,j+1} -\frac{i-2}{i-1}F_{i-2,j} 
\eqno(3.1)
$$

\textbf{when }$i\ge 3$\textbf{. }

This statement follows from 
formulas (1.1) and (2.1).

\subsection{\textbf{The matrix} $\bf F$\textbf{ is an upper triangular one, i.e. }
$F_{ij} =0$\textbf{, when }$i>j$\textbf{.}}

The polynomial $P_i \left( \eta \right)$ is orthogonal to 1, $\eta $, $P_2 
\left( \eta \right)$ when $i>2$. Hence $P_i \left( \eta \right)\bot \eta 
^2$. Continuing this argument we can conclude that if $P_i \left( \eta 
\right)$ is orthogonal to all $\eta ^k$ when $k<j$, then, because $P_i 
\left( \eta \right)\bot P_j \left( \eta \right)$, we have $P_i \left( \eta 
\right)\bot \eta ^j$. From here the proof of Property 4 follows.

\subsection{\textbf{The entries of the matrix} $\bf F$}

The entries of the first and second lines of the matrix $\bf F$ can be 
found from formula (\ref{eq5}): 

$$
F_{1,j} =\frac{2}{j} \quad {\rm if}\; j \;{\rm is \: odd,} \quad \quad
F_{2,j} =\frac{2}{j+1} \quad {\rm if}\; j \;{\rm is \: even.}
\eqno(5.1)
$$

An arbitrary entry of the matrix can be found using the Rodrigues formula 
for the Legendre polynomials
\[
P_n \left( \eta \right)=\frac{1}{2^nn!}\frac{d^n}{d\eta ^n}\left( {\eta 
^2-1} \right)^n.
\]
Then, in view of (1.1), 
\[
F_{ij} =\frac{1}{2^{i-1}\left( {i-1} \right)!}\sum\limits_{k=0}^{i-1} 
{\left( {-1} \right)^kC_{i-1}^k } \int\limits_{-1}^1 {\eta 
^{j-1}\frac{d^{i-1}}{d\eta ^{i-1}} \eta^{2\left( {i-k-1} 
\right)}} d\eta .
\]
From here it follows that

$$
F_{ij} =\frac{1}{2^{i-2}}\sum\limits_{k=0}^{\left[ {{\left( {i-1} \right)} 
\mathord{\left/ {\vphantom {{\left( {i-1} \right)} 2}} \right. 
\kern-\nulldelimiterspace} 2} \right]} {\left( {-1} \right)^k\frac{\left( 
{2i-2k-2} \right)!}{k!\left( {i-k-1} \right)!\left( {i-2k-1} \right)!\left( 
{i-2k-1+j} \right)}} .
\eqno(5.2)
$$
\newtheorem{lem}{Lemma}
\begin{lem}
\textbf{The diagonal entries of the matrix 
$\bf F$ are the following:
\[
F_{ii} =\frac{2^{i+1}i!\left( {i-1} \right)!}{\left( {2i} \right)!}.
\]}
\end{lem}
\textit{Proof. } Lemma is proved by induction. For $i=1$, this statement is obviously valid. 
We assume that it is true for $i=k$. From recurrence formula (3.1) and 
Property 4 it follows that
\[
F_{k+2,k} =\frac{2k+1}{k+1}F_{k+1,k+1} -\frac{k}{k+1}F_{kk} =0.
\]
Hence,
\[
F_{k+1,k+1} =\frac{k}{\left( {2k+1} \right)}\frac{2^{k+1}k!\left( {k-1} 
\right)!}{\left( {2k} \right)!}=\frac{2^{k+1}\left( {k!} \right)^2}{\left( 
{2k+1} \right)!}=\frac{2^{k+2}k!\left( {k+1} \right)!}{\left( {2k+2} 
\right)!} \quad .
\]
\begin{lem}
\textbf{The entries of the second upper diagonal 
of the matrix $\bf F$ are equal to
\[
F_{i-2,i} =\frac{2^{i-1}\left[ {\left( {i-1} \right)!} \right]^2}{\left( 
{2i-2} \right)!}\quad ,
\quad
i\ge 3.
\]}
\end{lem}
\textit{Proof}. Lemma is proved by induction. For $i=3$, this statement is valid (see 
(5.1)). Assume that it is true for $i=k$. Then from (3.1) and Lemma 1 it 
follows that 
\[
F_{kk} =\frac{2^{k+1}k!\left( {k-1} \right)!}{\left( {2k} 
\right)!}=\frac{2k-3}{k-1}F_{k-1,k+1} -\frac{k-2}{k-1}F_{k-2,k} .
\]
From here 
\[
F_{k-1,k+1} =\frac{2^{k+1}\left( {k-1} \right)k!\left( {k-1} 
\right)!}{\left( {2k-3} \right)\left( {2k} \right)!}+\frac{2^{k-1}\left( 
{k-2} \right)\left[ {\left( {k-1} \right)!} \right]^2}{\left( {2k-3} 
\right)\left( {2k-2} \right)!}=\frac{2^k\left( {k!} \right)^2}{\left( {2k} 
\right)!}\quad .
\]
Hence, the induction hypothesis also holds for $i=k+1$. 

\subsection{\textbf{Lemma about a linear combination of lines}}

\begin{lem}
\textbf{The vector ${\bf {f}'}=\left( {F_{1,m+1} ,F_{1,m+2} ,\ldots ,F_{1,m+n+1} 
} \right)^T$ is a 
linear combination of the vectors ${\bf f}_{\delta ,\bullet } ,{\bf f}_{\delta 
+2,\bullet } \ldots ,{\bf f}_{m+1,\bullet } $, 
 where $\delta =1$ if $m$ is even, and $\delta =2$ 
 if $m$ is odd.}
\end{lem}
Here we denoted ${\bf f}_{k, \bullet} = (F_{k1} ,F_{k2} ,\ldots ,F_{k,n+1})^T$ 
for any natural $k$.
 
\textit{Proof}.

We write down formula (5.2) in the form
$$
F_{ij} =\frac{1}{2^{i-1}}\sum\limits_{k=\delta ,\left( 2 \right)}^i {\left( 
{-1} \right)^{{\left( {i-k} \right)} \mathord{\left/ {\vphantom {{\left( 
{i-k} \right)} 2}} \right. \kern-\nulldelimiterspace} 2}\frac{\left( {i-k-2} 
\right)!}{\left( {k-1} \right)!\left( {{\left( {i-k} \right)} 
\mathord{\left/ {\vphantom {{\left( {i-k} \right)} 2}} \right. 
\kern-\nulldelimiterspace} 2} \right)!\left( {{\left( {i+k} \right)} 
\mathord{\left/ {\vphantom {{\left( {i+k} \right)} 2}} \right. 
\kern-\nulldelimiterspace} 2-1} \right)!}} \frac{2}{\left( {j+k-1} \right)}
\eqno(6.1)
$$
where $\delta =1$ if $i$ is odd, and $\delta 
=2$ if $i$ is even.

In formula (6.1) and further on, the index ``(\ref{eq2})'' in the lower part of the 
sum symbol means that summation is carried out with a step equal to 2. Then, 
taking into account (5.1), formula (6.1) can be written as
$$
F_{ij} =\sum\limits_{k=\delta ,\left( 2 \right)}^i {\beta _{ki} F_{1,j+k-1} 
} ,
\eqno(6.2)
$$
where
$$
\beta _{ki} =\frac{1}{2^{i-1}}\left( {-1} \right)^{{\left( {i-k} \right)} 
\mathord{\left/ {\vphantom {{\left( {i-k} \right)} 2}} \right. 
\kern-\nulldelimiterspace} 2}\frac{\left( {i+k-2} \right)!}{\left( {k-1} 
\right)!\left( {{\left( {i-k} \right)} \mathord{\left/ {\vphantom {{\left( 
{i-k} \right)} 2}} \right. \kern-\nulldelimiterspace} 2} \right)!\left( 
{{\left( {i+k} \right)} \mathord{\left/ {\vphantom {{\left( {i+k} \right)} 
2}} \right. \kern-\nulldelimiterspace} 2-1} \right)!}\quad .
\eqno(6.3)
$$
If the vector ${\bf {f}'}$ is a linear combination of the vectors ${\bf f}_{\delta 
,\bullet } ,{\bf f}_{\delta +2,\bullet } \ldots ,{\bf f}_{m+1,\bullet } $, then its 
$j$-th component must be equal to the $j$-th element of a linear combination 
of the vectors $\sum\limits_{k=\delta ,\left( 2 \right)}^{m+1} {\alpha _k 
{\bf f}_{k,\bullet } } $ :
$$
F_{1,j+m} =\sum\limits_{k=\delta ,\left( 2 \right)}^{m+1} {\alpha _k 
\sum\limits_{l=\delta ,\left( 2 \right)}^k {\beta _{lk} F_{1,j+l-1} 
=\sum\limits_{l=\delta ,\left( 2 \right)}^{m+1} {F_{1,j+l-1} 
\sum\limits_{k=l,\left( 2 \right)}^{m+1} {\alpha _k \beta _{lk} } } } } .
\eqno(6.4)
$$
We choose the coefficients of a linear combination $\alpha _\delta ,\alpha 
_{\delta +2} ,\ldots ,\alpha _{m+1} $ in such a way that equality (6.4) be 
valid for any values $j$, where $j=\delta ,\delta +2,\ldots ,n+1$. For the 
next values of $j$ in the interval from 1 up to $n+1$, equality (6.4) is 
automatically holds, because of the zero entries of the matrix (Property 2). 
Then the factors before the values $F_{1,j+l-1} $ in (6.4) must be equal to 
zero when $l=\delta ,\delta +2,\ldots ,m-1$ and $\alpha _{m+1} \beta _{m+1} 
=1$. Therefore, we obtain the following system of linear equations with a 
square matrix having the dimension $\left( {\frac{m+3-\delta }{2}} 
\right)\times \left( {\frac{m+3-\delta }{2}} \right)$:
$$
\left( {\begin{array}{l}
 1\quad \;\beta _{\delta ,\delta +2} \quad \beta _{\delta ,\delta +4} \quad 
\;\ldots \quad \beta _{\delta ,m+1} \\ 
 0\quad \beta _{\delta +2,\delta +2} \;\beta _{\delta +2,\delta +4} \; 
\ldots \quad \beta _{\delta +2,m+1} \\ 
 0\quad \quad 0\quad \quad \, \beta _{\delta +4,\delta +4} \, \; \ldots \quad 
\beta _{\delta +4,m+1} \\ 
 \ldots \;\quad \;\ldots \quad \quad \ldots \; \quad \;\ldots \quad 
\quad\ldots \\ 
 0\quad \;\;\;\ldots \quad \quad \ldots \; \;\quad \;\ldots \quad \beta 
_{m+1,m+1} \\ 
 \end{array}} \right)\quad \left( {\begin{array}{l}
 \alpha _\delta \\ 
 \alpha _{\delta +2} \\ 
 \alpha _{\delta +4} \\ 
 \;\;\vdots \\ 
 \alpha _{m+1} \\ 
 \end{array}} \right)=\left( {\begin{array}{l}
 0 \\ 
 0 \\ 
 \,\vdots \\ 
 0 \\ 
 1 \\ 
 \end{array}} \right)
\eqno(6.5)
$$
This is a system of equations with a triangular matrix which has nonzero 
entries on the diagonal; hence it has the unique solution. Consequently, 
$\alpha _\delta ,\alpha _{\delta +2} ,\ldots ,\alpha _{m+1} $ are the 
required coefficients of expansion of the vector ${\bf {f}'}$ to the vectors 
${\bf f}_{\delta ,\bullet } ,{\bf f}_{\delta +2,\bullet } \ldots ,
{\bf f}_{m+1,\bullet } $. 

\textbf{Corollary: }The system of equations (6.5) can be written down in a 
general form:
$$
\bf Ba=e\quad ,
\eqno(6.6)
$$
where ${\bf a}=\left( {\alpha _1 ,\alpha _2 ,\alpha _3 ,\ldots ,\alpha _{m+1} } 
\right)^T$, ${\bf e}=\left( {0,0,\ldots 0,1} \right)^T$, and the matrix $\bf B$ 
$$
{\mathbf B}=\left( {\begin{array}{l}
 1\quad 0\quad \beta _{13} \quad 0\quad \beta _{15} \quad 0\quad \ldots 
\quad \beta _{1,m+1} \\ 
 0\quad 1\quad \;0\quad \beta _{24} \quad 0\quad \beta _{26} \;\;\;\: \ldots 
\quad \beta _{2,m+1} \\ 
 0\quad 0\quad \beta _{33} \quad 0\quad \beta _{35} \;\;\;0\;\quad \;\ldots 
\quad \beta _{3,m+1} \\ 
 \ldots \;\ldots \; \ldots \; \: \ldots \;\; \ldots \;\; \ldots \quad 
\ldots \quad\quad \ldots \\ 
 0\quad 0\quad \; 0\quad \;\: 0\quad \;\;0\quad \;0\quad \ldots \quad \beta 
_{m+1,m+1} \\ 
 \end{array}} \right)\quad ,
\eqno(6.7)
$$
where $\beta _{ij} =0$ if $i+j$ is an odd number, and $m$ is an arbitrary 
natural number.

\subsection{\textbf{Lemma about orthogonality}}

\begin{lem}
\textbf{For the matrix $\bf F$ from 
(\ref{eq5}) and $\bf B$ from (6.7) when 
$m=n$, the following equality holds:
$$
\bf FB=D\quad ,
\eqno(7.1)
$$
where $\bf D$ is a diagonal matrix with 
diagonal entries equal to 
$$
D_{ii} =\frac{2}{2i-1}\quad .
\eqno(7.2)
$$}
\end{lem}
\textit{Proof}.

As the matrices $\bf F$ and $\bf B$ are the upper triangular ones, then 
their product will be the same matrix. We will prove that all the entries of 
the matrix $\bf D$ that are above the diagonal will be equal to zero. 
We multiply the $i$-th line of the matrix $\bf F$ by\textbf{ 
}the $j$\textbf{-}th row of the matrix $\bf B$, $i<j$. If $i+j$ is odd, 
then the product is equal to zero because of Property 2 and the same form of 
the matrix $\bf B$. Let $i+j$ be even. Then 
$$
D_{ij} =\left( {{\bf f}_{i,\bullet } ,{\mbox{\boldmath$\rm\beta$}}_{\bullet ,j} } 
\right)=\sum\limits_{l=0,\left( 2 \right)}^{j-i} {\beta _{i+l,j} F_{i,i+l} } 
.
$$
By virtue of (6.2)
\[
D_{ij} =\sum\limits_{l=0,(2)}^{j-i} {\beta _{i+l,j} \sum\limits_{k=\delta 
,\left( 2 \right)}^i {\beta _{ki} F_{1,i+l+k-1} =\sum\limits_{k=\delta 
,\left( 2 \right)}^i {\beta _{ki} \sum\limits_{l=0,\left( 2 \right)}^{j-i} 
{\beta _{i+l,j} F_{1,i+l+k-1} } } } } ,
\]
where $\delta =1$ if $i,j$ are odd, and $\delta =2$ if $i,j$ are even.

On the other hand, by virtue of Property 4, 
\[
F_{jk} =\sum\limits_{m=\delta ,\left( 2 \right)}^j {\beta _{mj} F_{1,k+m-1} 
=\sum\limits_{m=\delta ,\left( 2 \right)}^{i-2} {\beta _{mj} } F_{1,k+m-1} 
+\sum\limits_{l=0,\left( 2 \right)}^{j-i} {\beta _{i+l,j} F_{1,i+l+k-1} =0} 
} .
\]
Then
\[
\begin{array}{l}
 D_{ij} =-\sum\limits_{k=\delta ,\left( 2 \right)}^i {\beta _{ki} 
\sum\limits_{m=\delta ,\left( 2 \right)}^{i-2} {\beta _{mj} F_{1,k+m-1} } } 
= \\ 
 -\sum\limits_{m=\delta ,\left( 2 \right)}^{i-2} {\beta _{mj} 
\sum\limits_{k=\delta ,\left( 2 \right)}^i {\beta _{ki} F_{1,k+m-1} 
=-\sum\limits_{m=\delta ,\left( 2 \right)}^{i-2} {\beta _{mj} F_{im} =0} } } 
\\ 
 \end{array}\quad .
\]
The latter equality is carried out once again because of Property 4. Hence, 
the matrix $\bf D$ is a diagonal one. To find the values of 
diagonal entries of this matrix, we use Lemma 1 and formula (6.3). In this 
case
\[
D_{ii} =F_{ii} \beta _{ii} =\frac{2^{i+1}i!\left( {i-1} \right)!}{\left( 
{2i} \right)!}\frac{1}{2^{i-1}}\frac{\left( {2i-2} \right)!}{\left[ {\left( 
{i-1} \right)!} \right]^2}=\frac{2}{2i-1}\quad ,
\]
and Lemma is proved. 

\subsection{\textbf{The inverse matrix entries }}

The entries of the inverse matrix $\bf G=F^{-1}$ can be determined from equation 
(7.1) with allowance for (6.3) and (7.2). Really, $\bf G=BD^{-1}$, consequently,
$$
G_{ij} =\left( {-1} \right)^{{\left( {j-i} \right)} \mathord{\left/ 
{\vphantom {{\left( {j-i} \right)} 2}} \right. \kern-\nulldelimiterspace} 
2}\frac{1}{2^j}\frac{\left( {2j-1} \right)\left( {j+i-2} \right)!}{\left( 
{i-1} \right)!\left( {{\left( {j-i} \right)} \mathord{\left/ {\vphantom 
{{\left( {j-i} \right)} 2}} \right. \kern-\nulldelimiterspace} 2} 
\right)!\left( {{\left( {j+i} \right)} \mathord{\left/ {\vphantom {{\left( 
{j+i} \right)} 2}} \right. \kern-\nulldelimiterspace} 2-1} \right)!}\quad ,
\eqno(8.1)
$$
if $i+j$ is even, and $i\le j$. Otherwise $G_{ij} =0$.

\subsection{\textbf{The coefficients of expansion of the vector }${\bf {f}'}$
\textbf{ by lines of the matrix }$\bf F$}

The values of required coefficients $\alpha _1 ,\alpha _2 ,\alpha _3 ,\ldots 
,\alpha _{m+1} $ are determined from the solution to the system of linear 
equations (6.6). From (7.1) it follows that $\bf B=GD$. Then from $\bf GDa=e$ 
follows $\bf Da=Fe$ and, as $\bf Fe$ is the last row of the matrix $\bf F$ and 
the diagonal entries of the matrix $\bf D$ are determined in (7.2), 
then
$$
\alpha _i =\frac{2i-1}{2}F_{i,m+1} \quad .
\eqno(9.1)
$$

\end{appendix}

\bibliographystyle{my-h-elsevier}

\end{document}